\documentclass[aps,prb,twocolumn,showpacs]{revtex4-1}

\usepackage{amsfonts,amsbsy}
\usepackage{amsmath}
\usepackage{graphicx}

\begin{document}

\title{Tunneling Anisotropic Magnetoresistance of Helimagnet Tunnel Junctions}

\author{Chenglong Jia and Jamal Berakdar}

\affiliation{Institut f\"ur Physik, Martin-Luther Universit\"at Halle-Wittenberg, 06120 Halle(Saale), Germany}

\begin{abstract}
We theoretically investigate the angular and spin dependent transport in normal-metal/helical-multiferroic/ferromagnetic heterojunctions.
We find a tunneling anisotropic magnetoresistance (TAMR) effect due to the spiral magnetic order in the tunnel junction and to an effective spin-orbit coupling induced by the topology of the localized  magnetic moments in the multiferroic spacer.
 The predicted TAMR  effect is efficiently controllable  by an external electric field due to the magnetoelectric coupling.
\end{abstract}


\maketitle


\emph{Introduction.-} Transport across two  ferromagnetic  layers separated by a tunnel barrier depends in general on the relative orientation of the
 layers magnetizations \cite{Julliere}, giving rise to the
tunnel magnetoresistance (TMR) effect \cite{TMR}.
In the presence of spin-orbit interactions TMR  becomes  spatially anisotropic  \cite{GaMnAs,Fe-GaAs-Au,pTAMR}.
%
%
  Tunnel anisotropic magnetoresistance  TAMR is observed  not only in magnetic tunnel junctions
  with two ferromagnetic electrodes \cite{GaMnAs} but also in ferromagnetic/insulator/normal-metal systems such as Fe/GaAs/Au\cite{Fe-GaAs-Au}.
   Here we show that TAMR is a distinctive feature of normal-metal/multiferroic/ferromagnetic heterojunctions with the particular advantage of being
    electrically controllable.
     The coexistence of coupled electric and magnetic order parameters in multiferroics \cite{multiferroics}  holds the promise of new opportunities  for  device fabrications \cite{MFTJ,JB}. Our interest is focused on helimagnetic multiferroic   \cite{TbMnO3,RMnO3}. The topology of the local helical magnetic moments in these materials induces a resonant, momentum-dependent spin-orbit interaction \cite{JB}. The non-collinear magnetic order together with the induced spin-orbit coupling result in  uniaxial TAMR with a C$_{2v}$ symmetry. These two factors and their interplay determine the size and the sign of  TAMR. In particular, a linear dependence on the spiral helicity results in an electrically \cite{helicity} tunable spin-orbit interaction \cite{JB} by means of the magneto-electric coupling, and thus TAMR is electrically controllable  accordingly.

\begin{figure}[b]
\includegraphics[width=6.5cm,angle=-90]{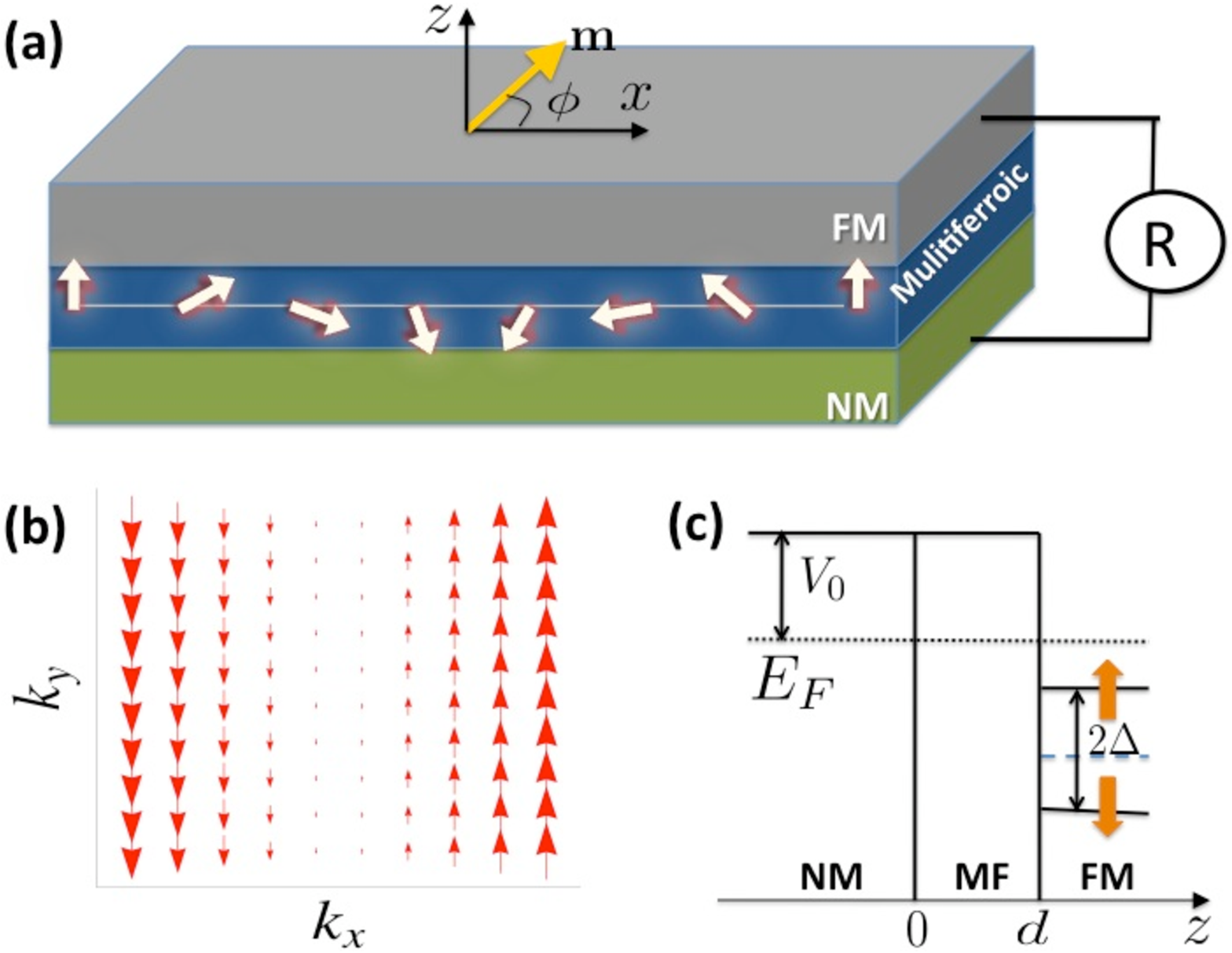}
\caption{(Color online) (a) Schematic diagram of the helical multiferroic tunnel junctions consisting of a normal metallic layer as the bottom electrode and a ferromagnetic layer for the top one. The vector $\mathbf{m}$ indicates the magnetization orientation specified by the angle $\phi$ in $xy$ (FM) plane. The $zx$ plane refers to the spiral plane of a multiferroic oxide. (b) The arrows show the induced resonant spin-orbit coupling, $qk_x \sigma_y$ in the multiferroic barrier. (c) The tunnel barrier potential profile.}
\label{fig::MTJ}
\end{figure}

\emph{Device proposition.-} The proposed multiferroic tunnel junction is sketched in Fig.\ref{fig::MTJ}. It consists of
 an ultrathin helical multiferroic barrier sandwiched between a normal metallic (NM) layer and a ferromagnetic conductor (FM). The ferroelectric polarization $\mathbf{P}$ in the multiferroic barrier  creates in general   surface charge densities $\pm |\mathbf{P}|$ which are screened by the induced  charge at the two metal electrodes \cite{TER}. A depolarizing field emerges in the barrier.  Taking the spontaneous electric polarization as $P_z = 700 \mu C/m^2$ and the dielectric constant to be $\epsilon = 30$ in the ferroelectric phase of TbMnO$_3$ \cite{TbMnO3}, the potential drop generated by the depolarizing field in the multiferroic barrier is estimated to be on the energy scale of $meV$, which is much smaller than any other relevant energy scale in the system. In the present study, we neglect this potential modification, and assume that the barrier potential has a rectangular shape  with the height $V_0$.  All energies
are given with respect to the NM Fermi energy $E_F$.
Based on this approximation, the Hamiltonians governing the carrier dynamics in
 the two electrodes and the oxide insulator have the following form,
\begin{eqnarray}
&& H_{NM} = - \frac{\hbar^2}{2m_e} \boldsymbol{\nabla}^2, ~~\mbox{for}~~ z<0, \nonumber\\
&& H_{MF} = -\frac{\hbar^2}{2m^*} \boldsymbol{\nabla}^2 + J \mathbf{n_r} \cdot \boldsymbol{\sigma} + V_0, ~~\mbox{for}~~ 0\leq z \leq d,\nonumber \\
&& H_{FM} = -\frac{\hbar^2}{2m_e} \boldsymbol{\nabla}^2 - \Delta \mathbf{m} \cdot \boldsymbol{\sigma} , ~~\mbox{for}~~ z>d,
\label{eq:1}\end{eqnarray}
where $V_0$ and $d$ are the height and the width of the potential barrier (see Fig.\ref{fig::MTJ}(c)), $m_e$ is the free-electron mass. $m^*$ is the effective electron mass of the oxide ( $m^*/m_e \approx 10$), and $\boldsymbol{\sigma}$ is the vector of Pauli matrices. $\mathbf{m} = [\cos \phi, \sin \phi, 0]$ is a unit vector defining the in-plane magnetization direction in the ferromagnet with respect to the [100] crystallographic direction. $\Delta$ is the half-width of the Zeeman splitting in the ferromagnetic electrode.   $J\mathbf{n_{r}}$ is the exchange field, where $\mathbf{n_{r}}$ is given by the multiferroic oxide local magnetization at each spiral layer (labelled by a integer number $l$) along the $z$-axis \cite{helicity}, i.e., $\mathbf{n_{r}} = (-1)^{l} [ \sin \theta_r, 0, \cos \theta_r]$ with $\theta_r = \mathbf{ \bar q}_m \cdot \mathbf{r}$ and $\mathbf{\bar q}_m = [\bar q,0,0]$ being the spiral spin-wave vector.
The physical picture behind the term $H_{MF}$ in eq.(\ref{eq:1}) is that a tunneling electron experiences an exchange coupling at the sites of the
 localized, non-collinear magnetic moments within the barrier. In effect this acts on the electron as a non-homogenous magnetic field.
 Performing a local unitary transformation within the barrier \cite{JB}, one can also view the influence of the barrier as consisting of two terms a homogeneous Zeeman field, and a topology-induced spin-orbit coupling SOC that depends solely on
  the helical magnetic ordering. As shown in \cite{JB}, this  SOC depends linearly  on the electron wave vector and on the helicity of the magnetic order \cite{JB} and is explicitly given by 
\begin{equation}  
\mbox{SOC} \sim \frac{\hbar^2}{2m^*} \bar q k_x \sigma_y. \end{equation}
The dependence on $k_x$ resembles the resonant semiconductor case
 when the Rashba \cite{Rashba} and Dresselhaus \cite{Dresselhaus} spin-orbit interactions have exactly equal strengths. Provided that the in-plane wave vector $\mathbf{k}_{\|}$ is no-zero, an electron in the oxide undergoes an exchange interaction with the local spiral magnetic moment and the induced spin-orbit coupling. So an electron spinor in the multiferroic barrier is determined by the following spin-dependence term,

\begin{equation}
H_{SO}^{eff} = \mathbf{w}(\theta_r,\mathbf{k}) \cdot \boldsymbol{\sigma}
\label{SOI}
\end{equation}
where 
\begin{equation}
\mathbf{w}(\theta_r,\mathbf{k}) = [(-1)^{l}J \sin \theta_r , q k_x, (-1)^{l}J \cos \theta_r]\end{equation} and $q= \frac{\hbar^2}{2m^*} \bar q.$ With this effective spin-orbit interaction we analyze the angular dependence of the electron tunneling through the helical multiferroic barrier.

\emph{Phenomenological theory.-}
We assume  the strength of the effective Zeeman field $|\mathbf{w}(\theta_r, \mathbf{k})|$ to be  relatively smaller than the Fermi energy $E_F$ and the band splitting $\Delta$.
Proceeding phenomenologically as in Ref.\onlinecite{pTAMR}  one  expands the transmissivity as a perturbative series of $\mathbf{m} \cdot \mathbf{w}(\theta_r,\mathbf{k})$. Up to the second order, the transmissivity reads,

\begin{eqnarray}
T(\mathbf{k},\mathbf{m}) &=& a_{1}^{(0)}(\mathbf{k}) + a_{1}^{(1)}(\mathbf{k}) [\mathbf{m} \cdot \mathbf{w}(\theta_r,\mathbf{k}) ] + a_{1}^{(2)}|\mathbf{w}(\theta_r,\mathbf{k})|^2 \nonumber \\ &+& a_{2}^{(2)}(\mathbf{k}) |\mathbf{m} \cdot \mathbf{w}(\theta_r,\mathbf{k})|^2
\end{eqnarray}
The expansion coefficients, $a_{i}^{(j)}$ ($i=1,2; ~ j=0,1,2$) satisfy the symmetry relations $a_{i}^{(j)}(k_x, k_y) = a_{i}^{(j)}(-k_x, -k_y)$, $a_{i}^{(j)}(k_x, k_y) = a_{i}^{(j)}(-k_x, k_y)$ and $a_{i}^{(j)}(k_x, k_y) = a_{i}^{(j)}(k_y, k_x)$. Based on the linear-response theory, the conductance $G(\phi)$ is found as

\begin{equation}
G(\phi) = \frac{e^2}{h} \int \frac{d^{2}\mathbf{k}_{\|}}{(2\pi)^2} \frac{d\theta_r}{2\pi} T(\mathbf{k},\mathbf{m}) =G_0 + G_{aniso}(\phi)
\end{equation}
where $G_0$ is the angular-independent part of the conductance, and

\begin{equation}
G_{aniso}(\phi) = g \, \mbox{tr}\, [A \, M(\phi)]
\label{aG}
\end{equation}
is the anisotropic spin-orbit coupling contributions and $g = e^2/ 8 \pi^3 h$. $A$ and $M(\phi)$ are matrices whose elements are given respectively by

\begin{equation}
A_{ij} = \langle a_{2}^{(2)}(\mathbf{k}) w_{i}w_{j} \rangle, ~~ M_{ij}(\phi) = m_{i} m_{j} ~~ (i,j=x,y,z).
\end{equation}
The notation $\langle ... \rangle$ stands for the integration over $\theta_r$ and $\mathbf{k}_{\|}$. Introducing  $\mathbf{w}(\theta_r,\mathbf{k})$ and $\mathbf{m}$ into $G_{aniso} (\phi)$, Eq.(\ref{aG}), the anisotropic conductance can be rewritten as

\begin{equation}
G_{aniso}(\phi) = \alpha \cos^2 \phi + \beta \sin^2 \phi.
\end{equation}
Considering the symmetry of the expansion coefficient $a_{2}^{(2)}(\mathbf{k})$ we obtain  for the above expression, $\alpha = g\langle a_{2}^{(2)}(\mathbf{k})J^2 \sin^2 \theta_r \rangle$ and $\beta = g\langle  a_{2}^{(2)}(\mathbf{k}) q^2 k_x^2 \rangle$.
Hence, the TAMR coefficient is given by

\begin{eqnarray}
\text{TAMR} =\frac{G(0)-G(\phi)}{G(\phi)}
\approx \gamma (1- \cos 2\phi),~ \gamma = \frac{\alpha - \beta}{2G_0}. 
\label{TAMR}
\end{eqnarray}
The above angular dependence of TAMR is quite general. The helical magnetic order and the induced spin-orbit interaction give rise to the anisotropy in the magentoconductance.
However, as evident from  Eq.(\ref{TAMR}), the TAMR coefficient $\gamma$ depends on $(\alpha - \beta)$, i.e.  contributions from the exchange interaction and the spin-orbit coupling term have opposite effects, which is confirmed by the following model calculations.

\emph{Ultrathin barriers.-}Experimental observations \cite{MFTJ} indicate that  thin film  multiferroics can retain both magnetic and ferroelectric properties down to a thickness of 2 nm (or even less). To get more insight
 in TAMR we consider  ultrathin tunneling barriers that can be approximated by a Dirac-delta function \cite{DDFM}.
 The effective spin-orbit interaction $H_{MF}^{\sigma}$ throughout the multiferroic barrier reduces then
 to the plane of the barrier, $\tilde{H}_{MF}^{\sigma} = \tilde{\mathbf{w}}(\theta_r,\mathbf{k}) \cdot \boldsymbol{\sigma} \delta(z)$ with $\tilde{\mathbf{w}}(\theta_r,\mathbf{k}) = [\tilde{J} \sin \theta_r, \tilde{q} k_x, \tilde{J} \cos \theta_r]$. $\tilde{J}$ and $\tilde{q}$ are renormalized exchange and resonant spin-orbit coupling parameters, $\tilde{q} \approx \bar{q}V_{0}d$ and $\tilde{J} \approx \langle J(z) \rangle_d$ referring to space and momentum averages with respect to the unperturbed states at the Fermi energy. In the following, we treat $\tilde{J}$ and $\tilde{q}$ as adjustable parameters. Obviously, the electron momentum parallel to the junction interfaces $\mathbf{k}_{\|}$ is conserved. Then the transverse electron wave functions in  NM ($z<0$) and FM ($z>0$) regions can be written as

\begin{eqnarray}
&& \Psi_{NM}^{\sigma} (z) = e^{i \kappa z} \chi_{\sigma} + r_{\sigma,\sigma} e^{-i \kappa z} \chi_{\sigma} + r_{\sigma, \tilde{\sigma}} e^{-i \kappa z} \chi_{\tilde{\sigma}},  \\
&& \Psi_{FM}^{\sigma} (z) = t_{\sigma,\sigma} e^{i k_{\sigma} z} \chi_{\sigma} + t_{\sigma,\tilde{\sigma}} e^{i k_{\tilde{\sigma}} z}  \chi_{\tilde{\sigma}}
\end{eqnarray}
with
\begin{eqnarray}
&&\kappa = \sqrt{ E/\frac{\hbar^2}{2m_e} -k_{\|}^2  } \\ && k_{\sigma} = \sqrt{ (E + \sigma \Delta)/\frac{\hbar^2}{2m_e} -k_{\|}^2  }
\end{eqnarray}
and the spinors introduced as
\begin{eqnarray}
\chi_{\sigma} = \frac{1}{\sqrt{2}}\left( \begin{array}{cc} 1 \\ \sigma e^{i \phi} \end{array} \right)
\end{eqnarray}
and correspond to an electron spin parallel ($\sigma =1$) or antiparallel ($\sigma = -1$) to the magnetization direction in the ferromagnetic electrode.
The reflection ($r_{\sigma,\sigma}$ and $r_{\sigma,\tilde{\sigma}}$) and transmission ($t_{\sigma,\sigma}$ and $t_{\sigma,\tilde{\sigma}}$) coefficients can be analytically obtained from the continuity conditions for $\Psi(z)$ and $\Psi^{\prime}(z)/m$ at $z=0$ \cite{DDFM},

\begin{eqnarray}
\Psi_{NM}^{\sigma}(0^{-}) &=& \Psi_{FM}^{\sigma}(0^{+}), \\
\frac{\hbar^2}{2m_e} \frac{d\Psi_{NM}^{\sigma}(z)}{dz} \Big |_{z=0^{-}} &+& (V_0 d + \tilde{\mathbf{w}} (\theta_r, \mathbf{k}) \cdot \boldsymbol{\sigma}) \Psi_{NM}^{\sigma}(0^{-}) \nonumber \\
&=& \frac{\hbar^2}{2m_e} \frac{d\Psi_{FM}^{\sigma}(z)}{dz} \Big |_{z=0^{+}} .
\end{eqnarray}
The transmissivity of a spin-$\sigma$ electron through the multiferric tunnel junctions reads
\begin{equation}
T_{\sigma}(E, \mathbf{k}_{\|},\theta) = \Re\left[\frac{k_{\sigma}}{\kappa} |t_{\sigma,\sigma}|^2 + \frac{k_{\tilde{\sigma}}}{\kappa} |t_{\sigma,\tilde{\sigma}}|^2 \right].
\end{equation}
For a small applied bias voltages, the conductance $G$ is determined by the states at the Fermi energy $E_F$ \cite{TER,DDFM,Book},
\begin{eqnarray}
G_{\sigma} = \frac{e^2}{h} \int \frac{d^{2}\mathbf{k}_{\|}}{(2\pi)^2} \frac{d\theta_r}{2\pi} T_{\sigma}(E_F,\mathbf{k}_{\|},\theta_r).
\end{eqnarray}
%
\begin{figure}[t]
\includegraphics[width=8cm]{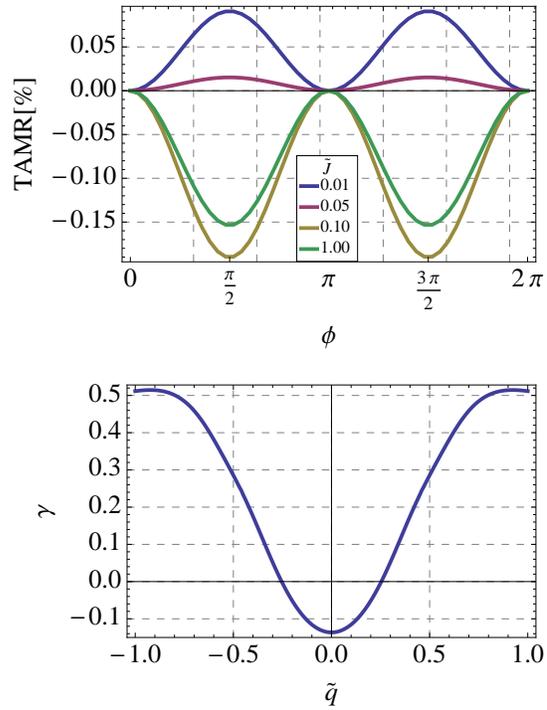}
\caption{The angular-dependence of TAMR on different strengths of the exchange field $\tilde{J}(eV)$ and on the coefficient $\gamma$  that
 enters the TAMR due to the competition between the exchange field and the induced spin-orbit interaction.
  The used numerical values are $E_F =5.5 eV$, $\Delta = 2 eV$, $V_0 = 0.5 eV$, and $d = 2$nm. Top: $\tilde{q} =0.28$. Bottom: $\tilde{J} = 0.1eV$.}
\label{fig::TAMR}
\end{figure}
%
Fig.\ref{fig::TAMR}(top) shows the TAMR angular dependence  $\sim (1- \cos 2\phi)$, at $E_F =5.5 eV$, $\Delta = 2 eV$, $V_0 = 0.5 eV$, and $d = 2$nm. It is clear that TAMR has $C_{2v}$ symmetry. For small $\tilde{J}$ the spin-orbit interaction dominates the tunneling properties, we have positive TAMR. As $\tilde{J}$ increases, the size of TAMR is influenced by the interplay between the exchange field and the induced spin-orbit interaction.  The TAMR is typically $\sim 0.1 \%$, which is on the same order as in the Fe/GaAs/Au tunnel junctions  have been recently
 realized   experimentally  \cite{Fe-GaAs-Au,YMnO3}. A transition from positive to negative TAMR is  observed, which is consistent with the previous finding, i.e. Eq.(\ref{TAMR}) within the phenomenological model. On the other hand, the helicity of the spiral magnetic order $\tilde{q}$ in helimagnetic multiferroics is experimentally controllable by a small ($\sim 1kV/cm$)  transverse electric field \cite{helicity}. Consequently,  we may electrically tune
 the spin-orbit coupling  strength and  thus TAMR in the helimagnet tunnel junctions (see Fig.\ref{fig::TAMR}(bottom)).

{\em Conclusions.-}  We  studied the electron tunneling properties through  helical multiferroic junctions. The spiral magnetic ordering and the induced spin-orbit interaction in the multiferroic barrier lead to the TAMR effect. Due to the magnetoelectric coupling, the strength of the induced spin-orbit coupling is electrically controllable which reders   a tunable TAMR by an external electric field.\\

This research is supported by the DFG (Germany) under SFB 762.


\end{document}